\documentclass[fleqn,usenatbib]{mnras}
\usepackage{anyfontsize}
\usepackage{newtxtext,newtxmath}
\usepackage[T1]{fontenc}
\DeclareRobustCommand{\VAN}[3]{#2}
\let\VANthebibliography\thebibliography
\def\thebibliography{\DeclareRobustCommand{\VAN}[3]{##3}\VANthebibliography}
\usepackage{graphicx}	
\usepackage{amsmath}	
\usepackage[dvipsnames]{xcolor}
\newcommand{\pcm}{\,cm$^{-3}$}
\title[NuSTAR QS flares]{Quiet Sun impulsive events observed with NuSTAR during solar minimum}
\author[S. Paterson et al.]{Sarah Paterson$^{1}$,
Iain G. Hannah$^{1}$\thanks{E-mail: iain.hannah@glasgow.ac.uk}, Brian W. Grefenstette$^{2}$, Säm Krucker$^{3,4}$, Erica Lastufka$^{5}$,
\newauthor{Hugh S. Hudson$^{1,3}$, Lindsay Glesener$^{6}$, Stephen M. White$^{7}$, David M. Smith$^{8}$}
\\
$^{1}$School of Physics \& Astronomy, University of Glasgow, University Avenue, Glasgow G12 8QQ, UK\\
$^{2}$Cahill Center for Astrophysics, California Institute of Technology, 1216 East California Boulevard, Pasadena, CA 91125, USA\\
$^{3}$Space Sciences Laboratory University of California, Berkeley, CA 94720, USA\\
$^{4}$University of Applied Sciences and Arts Northwestern Switzerland, 5210 Windisch, Switzerland\\
$^{5}$Department of Computer Science, University of Geneva, 7 route de Drize, 1227 Carouge, Switzerland\\ 
$^{6}$School of Physics \& Astronomy, University of Minnesota Twin Cities, Minneapolis, MN 55455, USA\\
$^{7}$Air Force Research Laboratory, Space Vehicles Directorate, Kirtland AFB, NM 87123, USA\\
$^{8}$Santa Cruz Institute of Particle Physics and Department of Physics, University of California, Santa Cruz, CA 95064, USA
}

\date{Accepted 2026 January 14. Received 2025 December 17; in original form 2025 July 28}

\pubyear{2026}

\begin{document}
\label{firstpage}
\pagerange{\pageref{firstpage}--\pageref{lastpage}}
\maketitle

\begin{abstract}
The investigation of small-scale energy release in the Sun's atmosphere is important in understanding how the corona is heated. Previous work has been able to study small EUV and SXR brightenings outside of active regions (i.e. the quiet Sun), but with HXRs this has mostly focused on active region transients/microflares due to the sensitivity of available telescopes. In this paper we present observations of the quiet Sun with the Nuclear Spectroscopic Telescope Array (NuSTAR), an X-ray imaging spectrometer with much greater sensitivity than previous instruments, allowing the observation of faint events. During the recent solar minimum, NuSTAR captured seven quiet Sun flares/impulsive brightenings, three on 21 February 2020, and four on 12--13 September 2020. From fitting their NuSTAR HXR spectra we find temperatures of 3.1--4.0~MK and emission measures between (0.75--17.0) $\times 10^{43}$~\pcm, which gives thermal energies between (2.5--8.9) $\times 10^{26}$~erg. Only one event, a mini-filament eruption, showed evidence of slightly higher temperatures emission, confirmed through Differential Emission Measure analysis. None of the events showed evidence of non-thermal emission in their NuSTAR spectra, and we placed upper limits to the  accelerated electron population. The thermal parameters for these quiet Sun events seem to scale differently to previously studied active region flares, suggesting a different energy release process might be dominating. However, this conclusion is affected by the different sensitivity and biases introduced by the various instruments and analysis approaches used.
\end{abstract}


\section{Introduction}

Investigating small-scale impulsive heating events in the quiet Sun is vital in understanding why the temperature of the solar corona is orders of magnitude higher than that of the Sun's surface. This phenomenon is termed the ``coronal heating problem'', and the heat source that is responsible for the observed high temperatures remains poorly understood. A possible solution was suggested by \citet{parker1988}, who proposed that the corona's high temperature could be sustained by an ensemble of tiny energy release events, termed ``nanoflares'', which would be too small to resolve individually.

Small-scale impulsive events are highly evident in both extreme ultraviolet (EUV) \citep{krucker1998,parnell2000,aschwanden2000,benz2002,2016A&A...591A.148J,madjarska2019,2022A&A...661A.149P} and soft X-rays (SXR) \citep{krucker1997,2021ApJ...912L..13V}. These studies found events down to energies of 10$^{24}$~erg and some of this work has focused on determining the index of the flare frequency distribution, which takes the form of a negative power law, with small-scale events dominating the distribution \citep{hudson1991}. While large flares do not release enough energy to explain the observed coronal heating, if the slope of the flare frequency distribution were > 2, then small-scale events would provide the dominant contribution to the heating. Recent work focused on nanoflares with SDO/AIA did find indices consistently > 2 \citep{2022A&A...661A.149P}. The results of previous observational studies have produced various indices ranging between 1.5--2.7 \citep{parnell2012} though this variation can arise from the different detection routines, assumptions and analysis approaches used \citep{2011SSRv..159..263H}.

These quiet Sun heating events typically show temperatures about 1~MK or lower in EUV, or slightly above this in SXR \citep{krucker1997}. The higher spatial and temporal resolution achieved by EUI on Solar Orbiter has resulted in large number of EUV brightenings being studied \citep{berghmans2021,2022A&A...663A.128A,2023A&A...676A..64N} but again finding events with temperatures predominantly at about 1~MK or lower \citep{2023A&A...671A..64D,2024A&A...688A..77D}. Some models suggest that evidence of higher temperature emission is required for small flares to heat the corona \citep{klimchuk2015}, so these EUV brightenings may not be a suitable candidate. Flares of higher temperature, from 2 to 7~MK, have been detected in SXR but from a non-imaging instrument \citep{2021ApJ...912L..13V}. No NOAA active regions were present during these times and EUV imaging observations were used to confirm the quiet Sun locations. Robustly classifying these events remains a major challenge given that they are at the limits of detectability - it is not clear what they are within the myriad of quiet Sun phenomena: microflares from weak active regions, flaring coronal/X-ray bright points, jets, mini-filament eruptions, network flares/brightenings, ``nanoflares''.

The energy release process in active region flares is observed to involve the acceleration of particles which then heats material into the corona. It is still unclear whether such a process is common in the small-scale energy releases in the quiet Sun. There is evidence of non-thermal emission from weak quiet Sun transients from radio observations \citep{2020ApJ...895L..39M,2022ApJ...937...99S} with one study suggesting the non-thermal energy in these events could be comparable or larger than the thermal energy \citep{2023ApJ...949...56M}. Hard X-ray (HXR) observations are ideal for studying the energetics of active region flares as they detect the thermal and non-thermal bremsstrahlung emission, finding that even the small microflares in active regions can still significantly accelerate electrons \citep{2008A&A...481L..45H,hannah2008,2024A&A...691A.172B}. However as solar HXR telescopes are optimised for active region events, earlier work was only able to produce upper limits to quiet Sun HXR emission \citep{hannah2007,hannah2010}. To study small-scale quiet Sun transients a HXR telescope with higher sensitivity is required and that opportunity arose with the Nuclear Spectroscopic Telescope Array (NuSTAR; \citet{harrison2013}). Though designed as an astrophysics mission, this HXR imaging spectrometer can also be used to observe the Sun and has helped to provide insight into small-scale energy release in the Sun's atmosphere. Much of the NuSTAR's solar analysis has focused on active region microflares \citep{wright2017,glesener2017,hannah2019,glesener2020,cooper2020,cooper2021,duncan2021,2024MNRAS.529..702C,bajnokova2024} finding high temperatures ($> 5$~MK) and non-thermal emission present in microflares orders of magnitude fainter than was previously studied. 

NuSTAR has also provided the unprecedented opportunity to detect extremely faint HXR sources in the quiet Sun. Quiet Sun flares observed with NuSTAR were first analysed by \citet{kuhar2018}. This work investigated three quiet Sun flares, observed on 2016 July 26 and 2017 March 21, away from the active regions present on the disk. The NuSTAR spectra of these three events were fitted to find temperatures and emission measures ranging between 3.2--4.1~MK and (0.6--15) $\times$ 10$^{44}$~\pcm, demonstrating that these quiet Sun flares were cooler and fainter than those found in NuSTAR active region microflare analysis, as well as no non-thermal emission being detected. Since the observations used in the analysis of \citet{kuhar2018}, there have been several NuSTAR quiet Sun observing campaigns, which took place during very quiet times over the last solar minimum (2018--2020). From September 2018 observations, a number of steady sources, such as X-ray bright points, were captured finding temperatures of 2-3~MK \citep{paterson2023}. A transient was detected in this work, a small jet, with the NuSTAR spectra consistent with a 2.6~MK source. Although no non-thermal emission was detected, upper limits placed on the possible non-thermal source showed it could still be sufficient to power the observed heating \citep{paterson2023}. Using NuSTAR observations from February and  September 2020, \citet{paterson2024} investigated the variability of X-ray bright points observed over several hours, finding that when they flared temperatures rose to just above 4~MK. Again no non-thermal emission was detected -- but modelling found that steep non-thermal spectra starting at a few keV could be present and provide enough energy to heat the thermal emission, whilst still being undetectable to NuSTAR.

\begin{figure*}
\includegraphics[width=\textwidth]{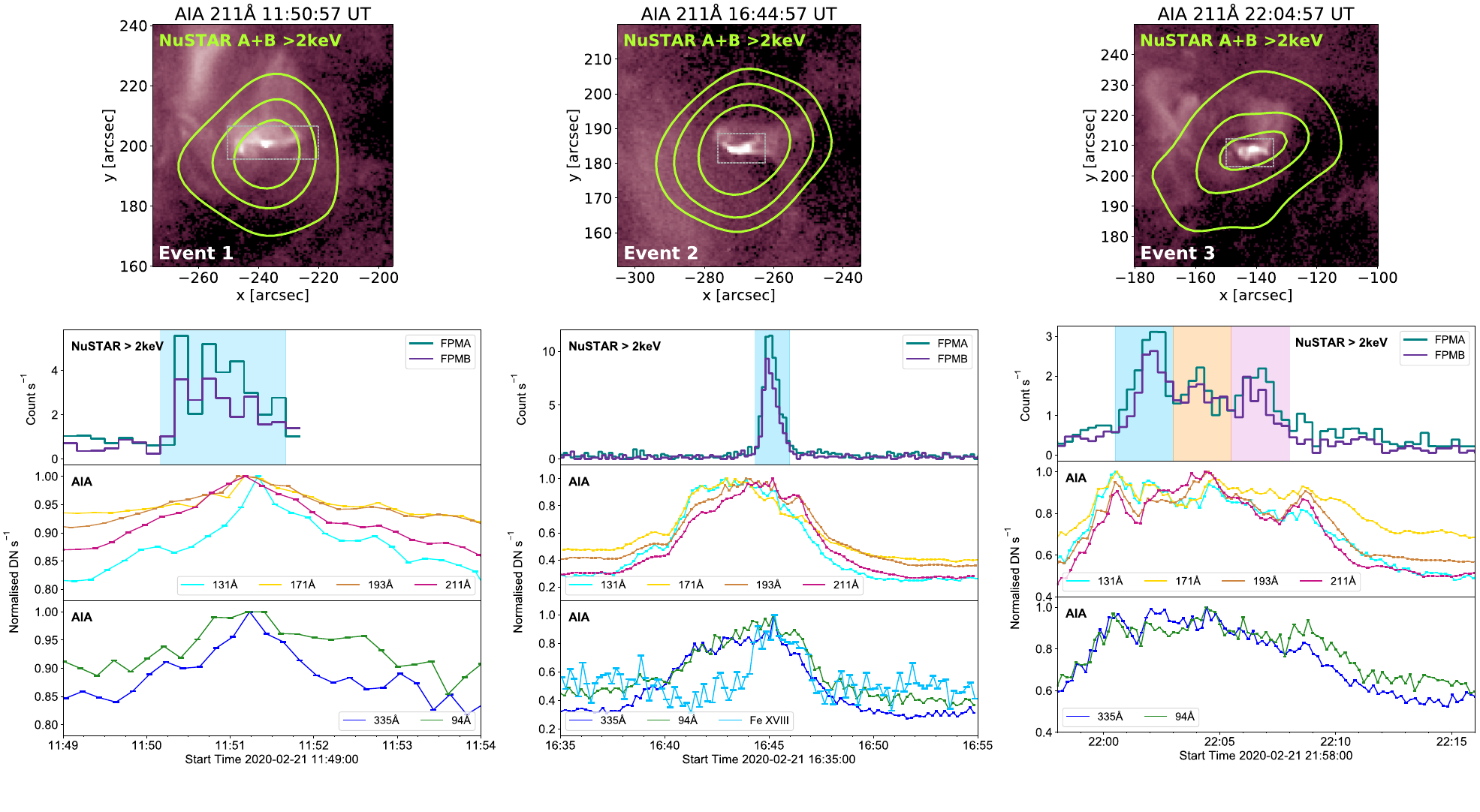}
    \caption{An overview of the three impulsive quiet Sun events (labelled Events 1--3) captured with NuSTAR on 21 February 2020. Top row: AIA 211\AA~ images of the flaring regions, with NuSTAR FPMA + FPMB contours shown in green. The AIA light curves were found over the grey dashed boxes. Bottom row: NuSTAR and AIA light curves showing the X-ray and EUV evolution of the events. Shaded regions represent the time intervals used for NuSTAR spectral fitting. }
    \label{fig:feboverview}
\end{figure*}
 
In addition to steady sources, NuSTAR also captured a number of transient HXR source during the solar minimum campaigns, similar to the impulsive quiet Sun energy release events studied by \citet{kuhar2018}. However, no active regions were present during these solar minimum campaigns and so NuSTAR's full throughput could be used to detect even fainter events. In this paper, we investigate the properties of seven events from these campaigns, observed on 21 February 2020 and 12--13 September 2020. These two observing campaigns are the longest dwells of NuSTAR on the very quiet Sun, which increased the chances of detecting short-lived transients. The evolution of longer duration X-ray bright points from these observations were studied in detail in \citet{paterson2024} - in this paper we focus on the short lived transients.  The detection of more transient events in addition to the three analysed in \citet{kuhar2018} allows a more thorough investigation into how the temperatures and emission measures of quiet Sun flares scale with those of active region microflares and flares.

We give an brief overview of the observing campaigns and of the each of the events (including images and time profiles) in Section~\ref{sec:overview}. In Section~\ref{sec:spec}, we present an overview of the NuSTAR spectral fitting for the events, as well as one in-depth example. In Section~\ref{sec:dem} the Differential Emission Measure (DEM) of the events is calculated. In Section~\ref{sec:thermalenergy}, we use the spectral fitting results to calculate the thermal energies of the seven events, and determine non-thermal upper-limits. Finally, we investigate how their temperatures and emission measures scale with those found for active region microflares and flares in previous studies in Section~\ref{sec:tem}.

\section{Overview of Events}
\label{sec:overview}

\begin{figure*}
\includegraphics[width=\textwidth]{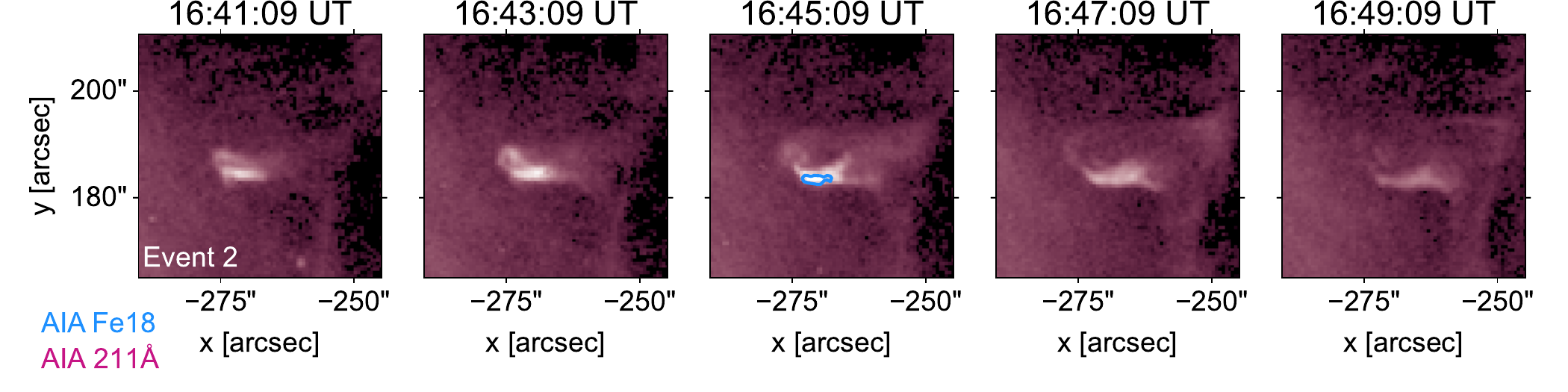}
    \caption{AIA 211\AA~ images showing the EUV evolution of the small mini-filament eruption captured by NuSTAR on 21 February 2020 (Event 2). The blue contours in the middle panel shows the 25\% level of the AIA Fe XVIII proxy channel.}
    \label{fig:mfimages}
\end{figure*}

\subsection{21 February 2020}
On 21 February 2020, NuSTAR observed the quiet Sun over eleven orbits (with each NuSTAR orbit having $\sim$ 1~hour in sunlight), with the first orbit beginning just after 05:15~UT and the last finishing at around 22:20~UT. The second and penultimate orbits were observed in full-disk mosaic mode, where NuSTAR's pointing was regularly changed to build up an image of the full disc. The other nine orbits were observed in dwell mode, where NuSTAR's pointing remained approximately constant (near disk centre). NuSTAR's livetime (the fraction of time that NuSTAR was able to detect incoming X-rays) during these orbits was 47--93$\%$, significantly higher than during microflare observations, when it is typically a fraction of a per cent \citep[e.g.][]{glesener2020}. While several persisting X-ray bright points were captured in this observation (the brightest of which is investigated in \citet{paterson2024}), a few transient events were also captured away from the bright points, manually identified via movies of each NuSTAR orbit.

\begin{figure*}
\includegraphics[width=0.7\textwidth]{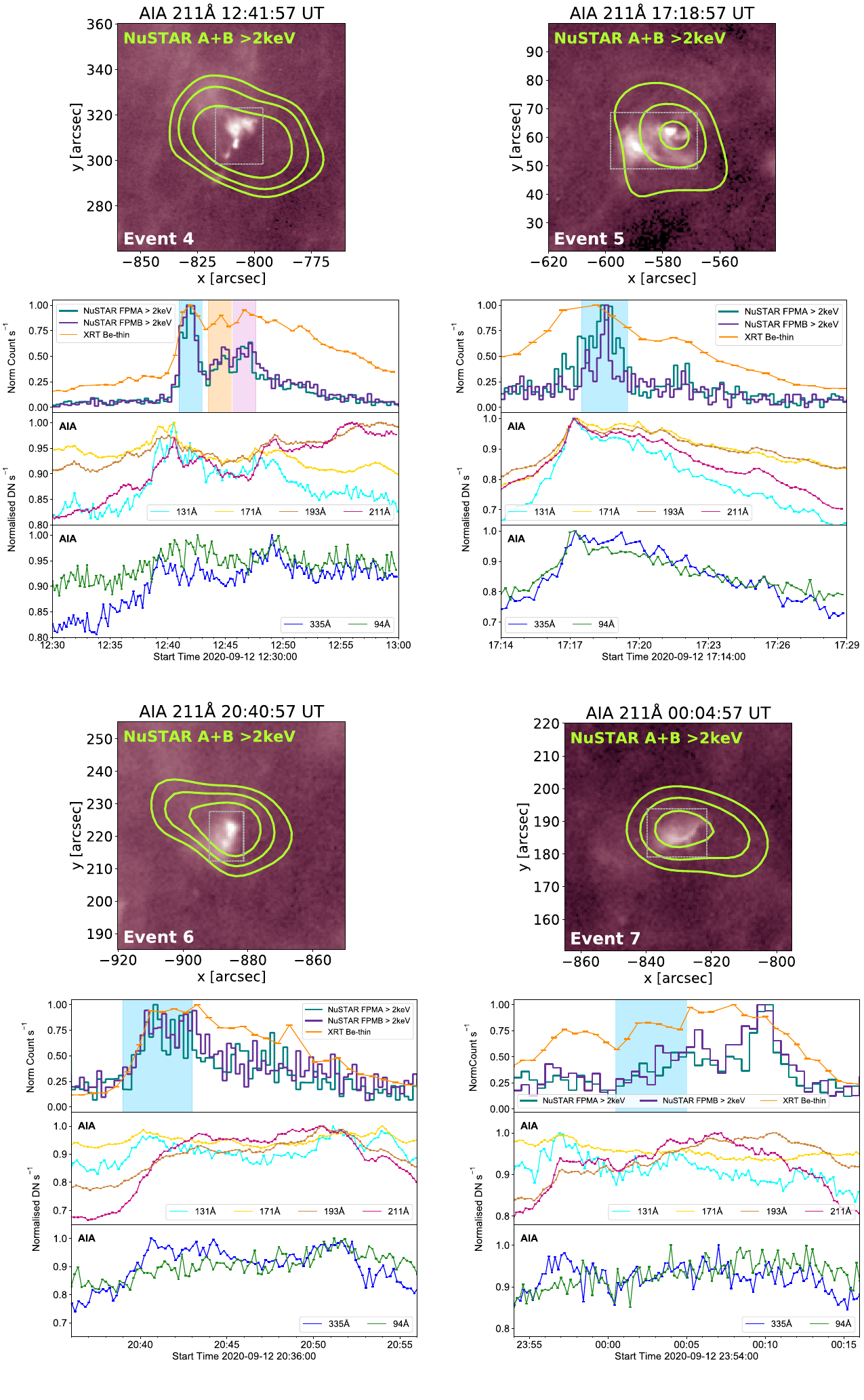}
    \caption{An overview of the four impulsive quiet Sun events (labelled Events 4--7) captured with NuSTAR on 12--13 September 2020. First and third rows: AIA 211\AA~ images of the flaring regions, with NuSTAR FPMA + FPMB contours shown in green. The AIA light curves were found over the grey dashed boxes. Second and fourth rows: NuSTAR, XRT, and AIA light curves showing the X-ray and EUV evolution of the events. Shaded regions represent the time intervals used for NuSTAR spectral fitting. Note that the main spike in the Event 7 time profile about 00:10 UT (bottom right panel) is due to a source outside the FOV.}
    \label{fig:sepoverview}
\end{figure*}

AIA and NuSTAR images and light curves for three quiet Sun transients captured on 21 February 2020 are shown in Fig.~\ref{fig:feboverview}. NuSTAR has two focal plane modules, FPMA and FPMB, and the contours shown on the AIA 211\AA~ images are summed over the two telescopes. These NuSTAR contours were shifted slightly, to align to the transient source in the 211\AA~ images per event. Note that we do not include any SXR observations from XRT for these events. This is because XRT was observing at a cadence of $\sim$ 6~minutes at the time of these brightenings, which is longer than the timescale on which these events evolved.

The NuSTAR time profile for Event 1 (left-hand panels in Fig.~\ref{fig:feboverview}) peaks at around 11:50~UT over about $\sim$ 2~minutes before any of the AIA channels, which is suggestive of plasma rapidly heating, before cooling out of the temperature range of NuSTAR's sensitivity and into a range to which the AIA channels are more sensitive. Event 3 (right-hand panels in Fig.~\ref{fig:feboverview}) has a more complicated evolution compared to the simple rise and decay of Event 1. The NuSTAR light curve for Event 3 exhibits  three distinct peaks, the strongest being the initial one at 22:02~UT. This event also has a longer duration than Event 1, only returning to background level after $\sim$ 8~minutes. The AIA and NuSTAR light curves for this event are not well-matched,and there is a steady increase in all the AIA channels, that reach a peak before the first NuSTAR peak. This could be due to an initial increase in emission at lower temperatures, only AIA is sensitive to \citep{lemen2012}. 

In Event 1 AIA images show small loops brighten then fade, and in Event 3 there is the hint of something heading outwards but confined. It is only in event 2 (middle panels in Fig.~\ref{fig:feboverview}) that clearly shows material being ejected, in the form of a mini-filament eruption. The EUV evolution of this event can be seen through the AIA 211\AA~ images shown in Fig.~\ref{fig:mfimages}, the time profiles plotted in Fig.~\ref{fig:feboverview} (bottom row, middle panel). Increased EUV emission is first detected by AIA at around 16:40~UT, and this slow rise continues until a peak is reached at 16:45~UT. At this time, there is a short (duration < 5~minutes), sharp spike in HXRs as observed by NuSTAR. It is at this point that material is seen to be ejected in AIA images, suggesting that the NuSTAR spike is due to X-ray emission from the heated flare loops/arcade underneath the erupting mini-filament, a configuration previously explored in other events in EUV and SXR \citep[c.f.][]{2015Natur.523..437S}. The decay in the AIA lightcurves is slower than that in NuSTAR, returning to background level by 16:50~UT. As demonstrated in Fig.~\ref{fig:feboverview} (bottom row, middle panel), 

This mini-filament eruption in Event 2, also produces a signature in the AIA Fe XVIII proxy channel, which is sensitive to plasma > 4~MK \citep{delzanna2013}. This is the only event of the ones studied here to be detectable in this channel. The time profile of the Fe XVIII emission is shown in Fig.~\ref{fig:feboverview} (bottom row, middle panel), with it peaking later than in the normal AIA channels, and more closely matched to NuSTAR, suggesting that they both observed plasma at similar temperatures. The image of the Fe XVIII emission about this peak time -- blue contours in the middle panel of Fig.~\ref{fig:mfimages} -- shows a small source co-spatial with the brightest 211\AA~ source.

\subsection{12--13 September 2020}
The second of the NuSTAR quiet Sun observing campaigns used in this analysis was carried out on 12--13 September 2020. This observation commenced at around 09:00~UT on 12 September and concluded the following day at around 00:35~UT. The first NuSTAR orbit was spent observing in mosaic mode. Then NuSTAR observed a region on the East limb in dwell mode from $\sim$ 10:30~UT onwards, for a total of nine orbits. During these orbits, NuSTAR's livetime was between 86--92$\%$. While an X-ray bright point emerged into the NuSTAR field-of-view (FOV) in the final three orbits of this observation (see \citet{paterson2024} for in-depth analysis of the evolution of this XBP), several short-lived sources were also captured.

Fig.~\ref{fig:sepoverview} shows AIA, NuSTAR, and XRT light curves for four transient X-ray brightenings in the quiet Sun captured by NuSTAR on 12--13 September, and NuSTAR contours aligned to the 211\AA~ AIA images. As XRT was observing at a higher cadence during this campaign, the SXR evolution of the events could be investigated in addition to the higher energy HXR emission observed by NuSTAR, and the results for the Be-Thin filter are shown.

It can be seen that the XRT and NuSTAR time profiles for these four events are generally well-matched. In Event 4 (top left panels in Fig.~\ref{fig:sepoverview}), the NuSTAR emission began to rise at around 12:40~UT, and returned to background level by 12:55~UT after a slow decay. The X-ray time profiles show multiple peaks --- including a strong initial peak at 12:42~UT. 

Event 5 (top right panels in Fig.~\ref{fig:sepoverview}) had a short duration in NuSTAR of $\sim$ 3~minutes, though a slower decay was observed in both AIA and XRT. AIA and XRT both observed the emission to peak earlier than NuSTAR, which contradicts the typical cooling scenario where the highest energy emission should peak first (as was seen in Event 1 in Fig.~\ref{fig:feboverview}, bottom left panel). This is likely a result of this event being caught on the edge of NuSTAR's FOV, meaning that NuSTAR missed some of the counts during the rise. 

Event 6 (bottom left panels in Fig.~\ref{fig:sepoverview}) shows a sharp rise and slow decay in NuSTAR, with a total duration of $\sim$ 10~minutes. The peak in NuSTAR coincides with peaks in the AIA 131 and 335\AA~ channels, and is followed a few minutes later by peaks in the 193 and 211\AA~ channels. This is again indicative of plasma cooling out of the range of NuSTAR's temperature sensitivity and into the range of these AIA channels.

Event 7 (bottom right panels in Fig.~\ref{fig:sepoverview}) is the faintest of the impulsive events analysed in this paper. The NuSTAR and XRT light curves are reasonably well-matched, though NuSTAR observed a peak at around 00:10~UT that was not observed by XRT. At this time, ghost rays (photons from sources outside the FOV \citep{madsen2015}) were detected across the whole NuSTAR detector. So, the highest spike in this NuSTAR time profile is therefore a result of another feature brightening and cannot be attributed to the event.

From the AIA images most of these events appear to be small loops, brightening to produce the flare/transients but it is challenging to get a clear structure even with AIA's spatial resolution. Event 4 does appear to be associated with a weak eruption, possibly a jet, but again it is difficult to determine this clearly. In the time profiles, the strong initial peak in the NuSTAR emission is similar to Event 2, which might again be suggestive of the hotter X-ray emission coming from heated flare loops/arcade underneath the erupting mini-filament. But Event 4 has additional X-ray emission (seen by both NuSTAR and XRT) after this initial peak, suggesting a more complicated post-flare structure.


\setlength{\tabcolsep}{0.5em} 
\renewcommand{\arraystretch}{1.2}
\begin{center}
    \begin{table*}
    \caption{A summary of the results from fitting all of the quiet Sun impulsive events' NuSTAR spectra with isothermal models, as well as the associated area from SDO/AIA, and calculated volume and thermal energy. Note that the events with more complex time profiles were split into multiple time intervals for this analysis. $\dagger$This event was also fitted by double thermal model, as shown in Fig.~\ref{fig:event2spec}.}
    \vspace{0.2cm}
    \centering
    \begin{tabular}{cccccccc}
    \hline
    Event&Date&Time Range&T&EM&211\AA~ Area&Volume&Thermal Energy\\
    &&[UT]&[MK]&[$\times$ 10$^{43}$ \pcm]&[arcsec$^{2}$]&$[\times$ 10$^{25}$ cm$^{3}$]&[$\times$ 10$^{25}$ erg]\\
    \hline
    1&2020-02-21&11:50:10--11:51:40&3.91{\raisebox{0.5ex}{\tiny$^{+0.18}_{-0.50}$}}&2.25{\raisebox{0.5ex}{\tiny$^{+2.41}_{-0.62}$}}&12&1.58
    &3.06{\raisebox{0.5ex}{\tiny$^{+0.78}_{-0.34}$}}\\
    \rule{0pt}{4ex}
    2$\dagger$ &2020-02-21&16:44:20--16:46:00&4.03{\raisebox{0.5ex}{\tiny$^{+0.06}_{-0.04}$}}&3.96{\raisebox{0.5ex}{\tiny$^{+0.49}_{-0.49}$}}&24&4.48
    &7.03{\raisebox{0.5ex}{\tiny$^{+0.35}_{-0.35}$}}\\
    \rule{0pt}{4ex}
    3&2020-02-21&22:00:30--22:03:00&3.60{\raisebox{0.5ex}{\tiny$^{+0.41}_{-0.25}$}}&1.98{\raisebox{0.5ex}{\tiny$^{+1.24}_{-0.83}$}}&28&5.65
    &4.98{\raisebox{0.5ex}{\tiny$^{+0.93}_{-0.75}$}}\\
     & &22:03:00--22:05:30&3.37{\raisebox{0.5ex}{\tiny$^{+0.40}_{-0.14}$}}&2.40{\raisebox{0.5ex}{\tiny$^{+1.17}_{-1.15}$}}& & \\
     & &22:05:30--22:08:00&3.05{\raisebox{0.5ex}{\tiny$^{+0.17}_{-0.39}$}}&5.23{\raisebox{0.5ex}{\tiny$^{+8.48}_{-1.73}$}}& & \\
    \rule{0pt}{4ex}
    4&2020-09-12&12:41:00--12:43:00&3.34{\raisebox{0.5ex}{\tiny$^{+0.11}_{-0.07}$}}&17.0{\raisebox{0.5ex}{\tiny$^{+3.70}_{-3.70}$}}&16&2.44
    &8.90{\raisebox{0.5ex}{\tiny$^{+0.72}_{-0.77}$}}\\
     & &12:43:30--12:45:30&3.22{\raisebox{0.5ex}{\tiny$^{+0.07}_{-0.04}$}}&13.0{\raisebox{0.5ex}{\tiny$^{+0.26}_{-2.60}$}}& & \\
     & &12:45:40--12:47:40&3.52{\raisebox{0.5ex}{\tiny$^{+0.35}_{-0.16}$}}&7.94{\raisebox{0.5ex}{\tiny$^{+2.97}_{-2.97}$}}& & \\
    \rule{0pt}{4ex}
    5&2020-09-12&17:17:30--17:19:30&3.41{\raisebox{0.5ex}{\tiny$^{+0.38}_{-0.14}$}}&7.80{\raisebox{0.5ex}{\tiny$^{+3.40}_{-3.47}$}}&20&3.41
    &7.28{\raisebox{0.5ex}{\tiny$^{+1.09}_{-1.25}$}}\\
    \rule{0pt}{4ex}
    6&2020-09-12&20:39:00--20:43:00&3.26{\raisebox{0.5ex}{\tiny$^{+0.13}_{-0.07}$}}&2.15{\raisebox{0.5ex}{\tiny$^{+0.65}_{-0.65}$}}&12&1.58
    &2.49{\raisebox{0.5ex}{\tiny$^{+0.29}_{-0.33}$}}\\
    \rule{0pt}{4ex}
    7&2020-09-13&00:00:30--00:05:00&3.35{\raisebox{0.5ex}{\tiny$^{+0.41}_{-0.13}$}}&0.75{\raisebox{0.5ex}{\tiny$^{+0.38}_{-0.37}$}}&40&9.64
    &3.73{\raisebox{0.5ex}{\tiny$^{+0.67}_{-0.75}$}}\\
    \hline
    \label{tab:fitresults}
    \end{tabular}
    \end{table*}
\end{center}

\section{NuSTAR Spectral Analysis}
\label{sec:spec}

In order to investigate the thermal properties of the seven events, we fitted their NuSTAR HXR spectra. We performed this analysis using XSPEC \citep{arnaud1996}, a program for X-ray spectral fitting, with the APEC thermal model, assuming solar coronal abundances \citep{1992ApJS...81..387F}. The faintness and short duration of these events can produce spectra which are noisy and difficult to fit, though this can be improved by simultaneously fitting the spectra from FPMA and FPMB, using a scaling factor as a fit parameter to account for systematic differences between the telescopes. Our fits are also improved by fitting down to energies of 2.2~keV instead of 2.5~keV, as is generally recommended for NuSTAR solar analysis \citep{grefenstette2016}. This approach is made possible by the NuSTAR calibration update detailed in \cite{madsen2021}, and has been used in previous analyses of the NuSTAR quiet Sun data from solar minimum \citep{paterson2023,paterson2024}. 

For each of the seven events, we chose the time of the strongest HXR emission for spectral fitting, all of which are summarised in Table~\ref{tab:fitresults}. These time ranges are indicated as blue shaded regions on the lightcurve plots shown in Figs.~\ref{fig:feboverview} and \ref{fig:sepoverview}. Four of these events (Events 1, 2, 5, and 6) exhibited simple X-ray time profiles, and we therefore considered only a single time interval. We also used one time interval for Event 7, which was chosen such that it did not coincide with any ghost rays from a brighter source outside the FOV or with an changes in NuSTAR's pointing (which occurred just before the beginning of the ghost rays). The X-ray lightcurves for both Events 3 and 4 showed three distinct spikes, and we fitted the NuSTAR spectra for these times separately. When fitting NuSTAR microflare spectra, a pre-flare thermal component would be used to take account of the active region emission, i.e. the background. This however is not required here as there was no significant bright pre-flare sources visible to NuSTAR, so there is no pre-flare background that needs to be subtracted. This is consistent with just the quiet corona being present before these events.

\begin{figure}
\includegraphics[width=1.0\columnwidth]{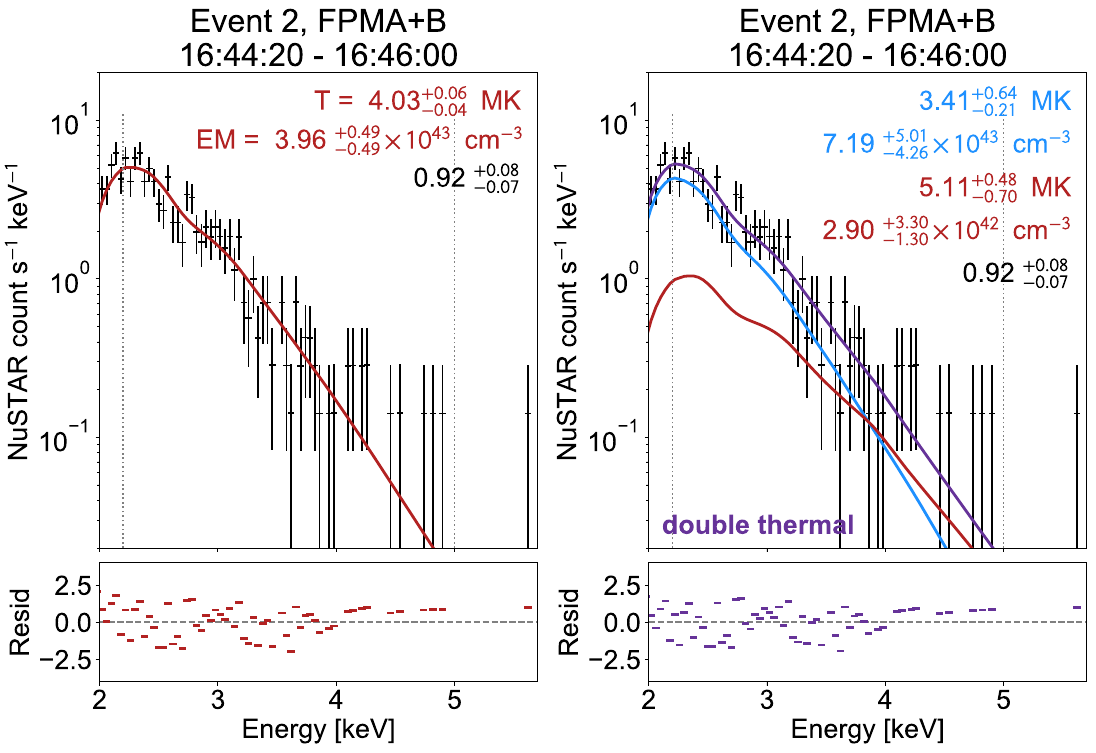}
    \caption{The NuSTAR FPMA+B spectrum for the mini-filament eruption. Left: the spectrum is fitted with an isothermal model (red). Right: the spectrum is fitted with a double thermal model (purple; the two separate thermal components are shown in red and blue). Temperatures and emission measures (as well as the multiplicative factor between FPMA and FPMB) as noted on the plots.}
    \label{fig:event2spec}
\end{figure}

An example of a fitted NuSTAR spectrum for one of the impulsive events, is shown in Fig.~\ref{fig:event2spec}. This example is for Event 2 (the mini-filament eruption), and we show this spectrum fitted with both an isothermal and a double thermal model, using the APEC thermal model, assuming solar coronal abundances \citep{1992ApJS...81..387F}. The isothermal fit gives a temperature of 4.03~MK and an emission measure of 3.96 $\times 10^{43}$~\pcm. However, there is a slight excess in the observed spectrum compared to the model at energies > 4~keV, which could indicate the presence of hotter emission. Due to this, and the earlier result that this mini-filament eruption produced a signature in AIA Fe XVIII (see Fig.~\ref{fig:feboverview}, bottom middle panel), indicating material heating to temperatures > 4~MK, we also tried fitting this spectrum with a double thermal model. The resulting fit gave a temperature of 3.41~MK and an emission measure of 7.19 $\times 10^{43}$~\pcm$\;$for the cooler component, and a temperature and emission measure of 5.11~MK and 2.90 $\times 10^{42}$~\pcm$\;$for the hotter component. It can be seen that both the isothermal and the double thermal models provide similar fits to the data, making it difficult to determine which is the better fit. The C-stat \citep{1979ApJ...228..939C} values are 144.3 for the 3 parameter isothermal fit, 143.8 for the double thermal 5 (2 additional) parameter fit, which does suggest there is little evidence the second component is statistically significant. We look more closely at the multi-thermal emission in Section~\ref{sec:dem}. Given the steepness of the spectra we do not believe a non-thermal model component could be robustly fitted to this data, and would be inconsistent with the signature in AIA Fe XVIII.

For all the other events only an isothermal model was required to well fit the NuSTAR HXR spectra, and the results of these fits (over the shaded time intervals in Figs. \ref{fig:feboverview} and \ref{fig:sepoverview}) are summarised in Table~\ref{tab:fitresults}. We found that the temperatures obtained from the NuSTAR spectral fits of the seven events (including the isothermal fit to Event 2) all lie within the narrow range of 3.1--4.0~MK, with emission measures varying between (0.75--17) $\times 10^{43}$~\pcm.

For Event 3, which had a NuSTAR time profile with three peaks (see Fig.~\ref{fig:feboverview}, bottom right panel), the NuSTAR spectral fits suggest plasma cooling, with the temperature decreasing (and emission measure increasing) successively with each peak between 3.60 and 3.05~MK. This is different to the case of Event 4, which also exhibited three distinct NuSTAR peaks. In this event, the temperature dropped from 3.34 to 3.22~MK between the first and second peaks, and then rose to 3.52~MK during the third, with the emission measure also falling during this final time period, possibly indicating additional energy release to heat the material.

The lowest emission measure, of 0.75 $\times 10^{43}$~\pcm, was for Event 7, the faintest of the events studied. However, it should be noted that ghost rays and NuSTAR pointing shifts made it impossible to fit the spectrum of this event during the time of peak X-ray emission as observed by XRT. Instead, a time interval during the early rise had to be used. Using the time of peak emission would likely result in a higher, though still very small, emission measure.

\section{Differential Emission Measure Analysis}
\label{sec:dem}
\begin{figure}
\includegraphics[width=.9\columnwidth]{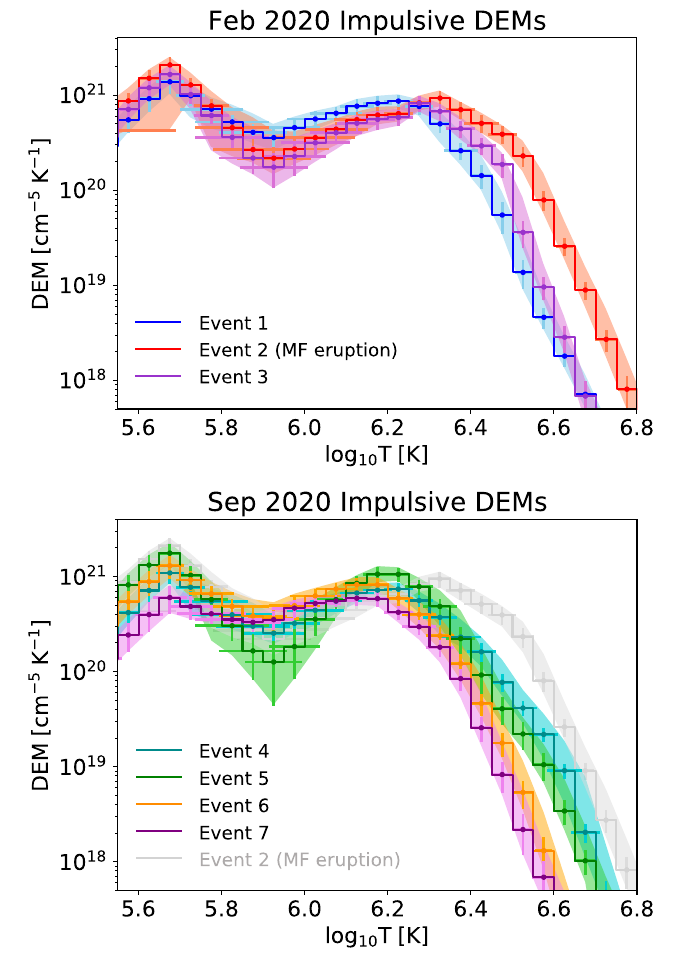}
    \caption{DEMs of impulsive events from February 2020 (top panel) and September 2020 (bottom panel), for the time ranges given in Table~\ref{tab:fitresults} - when more than one time is given the first is used - and shown by the blue shaded regions in Fig.~\ref{fig:feboverview} and Fig.~\ref{fig:sepoverview}. Note that Event 2 from February 2020 is also shown in the September 2020 panel, greyed out, for comparison.}
    \label{fig:dems}
\end{figure}

In order to investigate the multi-thermal emission in these events, in particular the presence of emission > 4MK in Event 2, we determine the Differential Emission Measure (DEM) for each of the events, using the approach of \citet{hannah2012}. For this we use the six coronal channels from AIA, the NuSTAR data split into two energy bands (2.2-3.2 keV and 3.2 -5 keV) and for the September 2020 event the XRT Be-thin (which is not available February 2020 events). This was calculated for the same areas used to produce the time profiles in Fig.~\ref{fig:feboverview} and Fig.~\ref{fig:sepoverview}, and the blue shaded regions in those figures are the time ranges used, which were also given in Table~\ref{tab:fitresults}. Uncertainties of the photon noise combined in quadrature with a systematic uncertainty of 20\% were used in the DEM calculation, a similar approach to those used previous \citep[e.g.][]{wright2017,paterson2024}. The temperature responses have all been calculated using CHIANTI version 9.3. The resulting DEMs for the 7 events are shown in Fig.~\ref{fig:dems}

The DEMs for all 7 events show a two peak structure, with one peak about $\log T=5.7$ from the background atmosphere, and another peak between $\log T=6.2-6.3$ which sharply fall off with increasing temperature. Out of all of these DEMs, Event 2's DEM (the mini-filament eruption) peaks at a slightly higher temperature and has the most material $\log T >6.4$, which is consistent with the findings from the spectral analysis in Section \ref{sec:spec}, confirming the presence of material heated to $> 4$ MK. For the other events there is a general consistency between the DEM and spectral results, for instance the smallest DEM and the weakest NuSTAR spectrum are both from Event 7.

\section{Energetics}
\subsection{Thermal Energy}
\label{sec:thermalenergy}

We used the parameters determined from the NuSTAR spectral fitting to estimate the thermal energy of the quiet Sun impulsive events. The thermal energy $U_\mathrm{T}$ of the events were calculated using \citet{hannah2008}, i.e.
\begin{equation}
    U_\mathrm{T} = 3k_\mathrm{B}T \sqrt{EM\times V}=3k_\mathrm{B}T \sqrt{EM\times A^{3/2}}
\end{equation}
\noindent where $k_\mathrm{B}$ is the Boltzmann constant, $T$ and $EM$ are the temperature and emission measure from the NuSTAR spectral fits and $V$ is the volume of the emitting plasma found from the area $A$ of the brightening source as seen in AIA 211\AA~ images, since this filter's temperature response is the closest to the temperatures found with NuSTAR (though slightly lower). The areas chosen where restricted to a small brightening region in the AIA images at the same time as the NuSTAR peak -- the regions used are shown in the left columns of Fig.~\ref{fig:nt_upper_feb} and \ref{fig:nt_upper_sep}, and given in Table~\ref{tab:fitresults}. We assumed a filling factor of 1, meaning that the calculated thermal energy is an upper limit.

The calculated thermal energies of the seven impulsive events are listed in Table~\ref{tab:fitresults}. These were found to range between (2.5--8.9) $\times 10^{25}$ erg. The thermal energies of the events presented in \cite{kuhar2018} were approximately an order of magnitude higher than these, lying between (2--6) $\times 10^{26}$~erg. This is likely a result of the fainter events being detected in the work presented here arising from NuSTAR's higher livetime and the lower background during solar minimum.

\subsection{Non-Thermal Energy}

\begin{figure*}
\includegraphics[width=1.3\columnwidth]{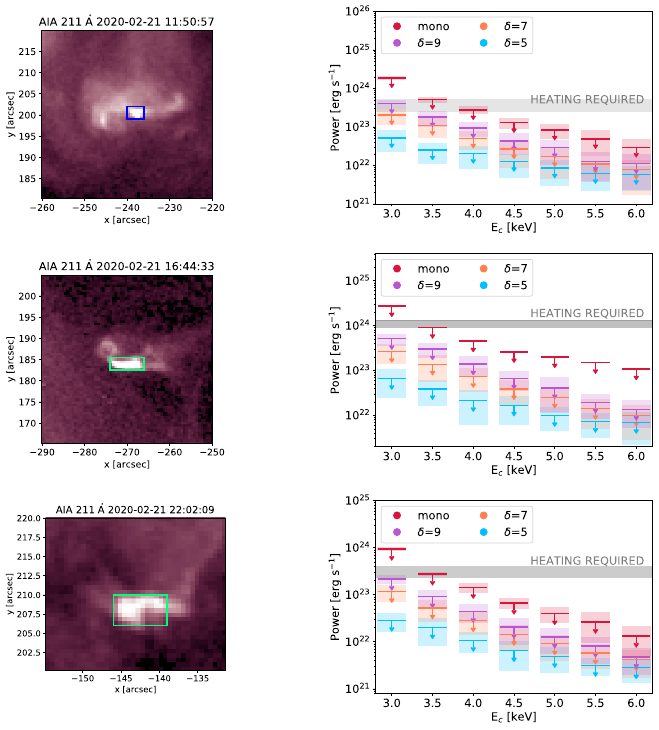}
    \caption{Non-thermal upper limits for Events 1-3 (top to bottom row) from February 2020. (Left) AIA 211~\AA images showing the area used for the calculation of the volume and thermal energy. (Right) Upper limits on the non-thermal emission for each event for a range of power law indices. The coloured shaded regions indicate the $\pm1\sigma$ range, and the grey line indicates the heating requirement dictated by the thermal energy.\label{fig:nt_upper_feb}}
\end{figure*}

\begin{figure*}
\includegraphics[width=1.3\columnwidth]{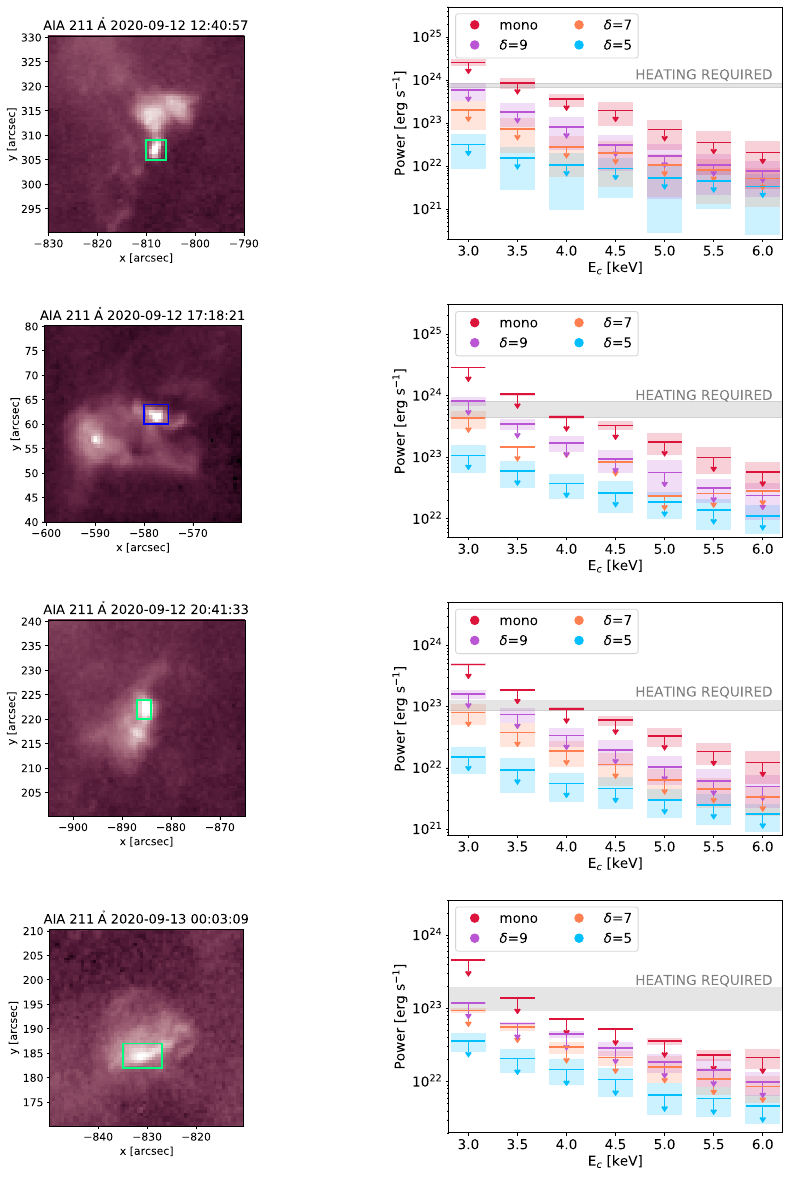}
    \caption{Non-thermal upper limits for Events 4-7 (top to bottom row) from September 2020. (Left) AIA 211~\AA images showing the area used for the calculation of the volume and thermal energy. (Right) Upper limits on the non-thermal emission for each event for a range of power law indices. The coloured shaded regions indicate the $\pm1\sigma$ range, and the grey line indicates the heating requirement dictated by the thermal energy.\label{fig:nt_upper_sep}}
\end{figure*}

The NuSTAR spectra in all of the events showed no evidence of a non-thermal component, as discussed in Section~\ref{sec:spec}. However these spectra are noisy or have no counts at higher energies (which could be due to NuSTAR's limited detector throughput) and a weak non-thermal component could be present but undetected. We can determine an upper limit on the non-thermal emission that could be present, remain consistent with a null detection and contain enough energy to power the observed heating using the approach detailed in \citet{wright2017}, \citet{cooper2020} and \citet{paterson2023}. This is achieved by adding a non-thermal model component to the thermal model found from the NuSTAR spectral fitting. This non-thermal model used is the cold thick-target model of single power-law distribution of electrons which depends on the power-law index, $\delta$, the low-energy cutoff, $E_{c}$, and the total electron flux, $N_{N}$. For a chosen $\delta$, $E_{c}$ and $N_{N}$ value, the resulting non-thermal model and the fitted thermal model are folded through the NuSTAR response used in the spectral fitting of each event to produce a model count spectra. These simulations were done in Python using the thermal and thick-target models from the sunkit-spex package\footnote{\url{https://github.com/sunpy/sunkit-spex/blob/main/sunkit_spex/legacy/fitting/photon_models_for_fitting.py}}.From these, synthetic spectra are generated through a Monte Carlo process, randomly sampling the model count spectra for the total number of counts (calculated using the livetime and duration of the observation). For a range of different $\delta$ and $E_{c}$ combinations, $N_{N}$ is reduced until these synthetic spectra are within the Poisson uncertainty between 2 and 4 keV of the observed spectra, and there are $<$ 4 counts above 4-5 keV (depending on the event) -- consistent with a null detection to 2$\sigma$ \citep{1986ApJ...303..336G}. Because the observed spectra considered here are noisy, this process was repeated multiple (1000) times to provide an uncertainty range for each upper limit. As well as testing models with different $\delta$ values, the case of a mono-energetic beam of electrons, with an energy of $E_{c}$, can also be tested. The resulting power of these non-thermal components can then be calculated via
\begin{equation}
P(>E_{c}) = 1.6 \times 10^{-9}\left(\frac{\delta-1}{\delta-2}\right) N_{N}E_{c}.
\label{eq:power}
\end{equation}
The results of these simulations are shown as a function of $E_{c}$ and $\delta$ in the right column of Fig.~\ref{fig:nt_upper_feb} for Events 1-3 and \ref{fig:nt_upper_sep} for Events 4-7. These are shown in comparison to the heating requirement for each event, which is calculated as the thermal energy divided by the duration, the values given in Table~\ref{tab:fitresults}, finding $1-7\times 10^{23}\;\mathrm{erg\;s}^{-1}$. For all 7 events only a very steep or mono-energetic non-thermal population with $E_{c}=3-3.5$~keV would be consistent with the heating requirements and null detection. Even though our thermal energy, and hence heating requirements, are upper limits, it seems that if accelerated electrons are present in these events, they do not contain enough energy to power the observed heating. 

\section{EM vs T Comparison With Previous Studies}
\label{sec:tem}

\begin{figure*}
\includegraphics[width=1.3\columnwidth]{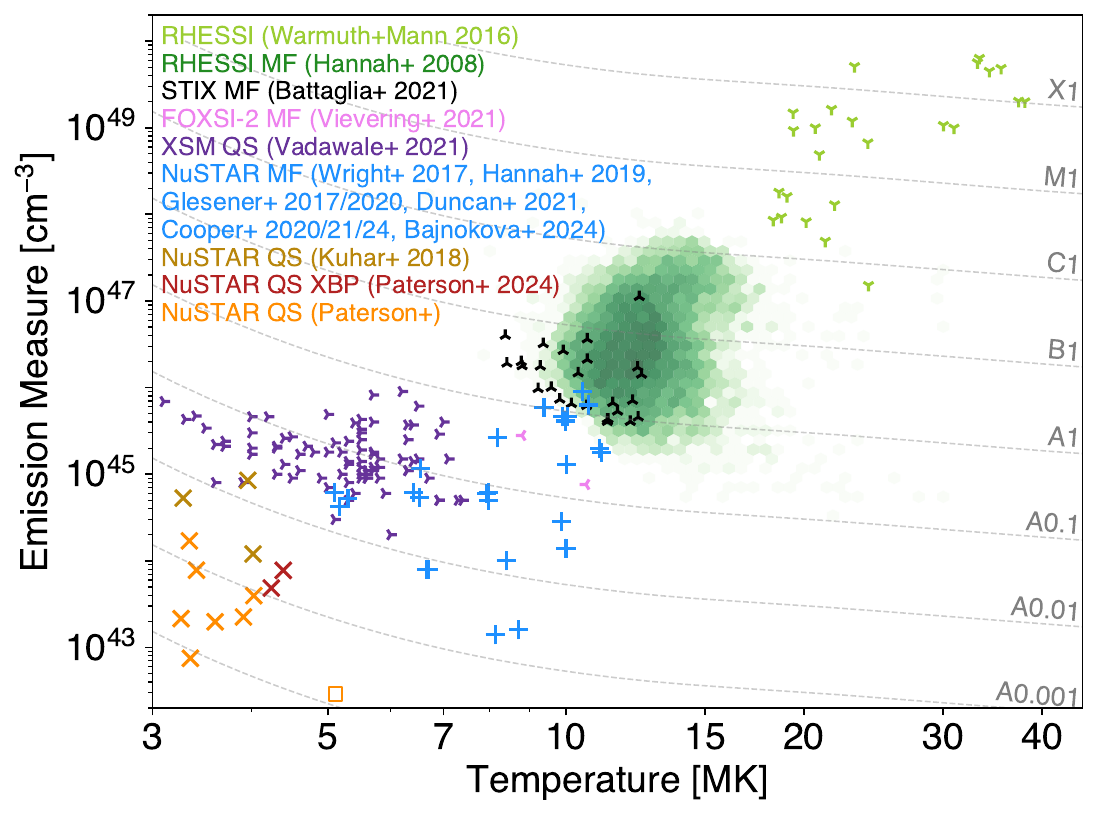}
    \caption{Isothermal emission measure against temperature for flares of various sizes from active regions and the quiet Sun, found through fitting their X-ray spectra. The results for the NuSTAR quiet Sun impulsive events from this study are shown as orange crosses, the orange square being the higher temperature found for Event 2. Included are previous NuSTAR studies of active region microflares \citep[blue pluses -][]{wright2017,hannah2019,glesener2017,glesener2020,duncan2021,cooper2020,cooper2021,2024MNRAS.529..702C,bajnokova2024}, quiet Sun X-ray bright point flares \citep[dark red crosses -][]{paterson2024} and quiet Sun impulsive events/flares \citep[dark yellow crosses -][]{kuhar2018}. Also included are previous studies of large flares observed with RHESSI \citep[green tri down -][]{warmuth2016}, active region microflares observed with RHESSI \citep[green hex bins -][]{hannah2008}, the FOXSI-2 sounding rocket \citep[pink tri left -][]{vievering2021}, STIX on Solar Orbiter \citep[black tri up -][]{battaglia2021}, and quiet Sun flares observed with Chandrayaan-2 XSM \citep[purple tri right -][]{2021ApJ...912L..13V}. Grey dashed lines indicate GOES class as a function of temperature and emission measure.}
    \label{fig:templot}
\end{figure*}

The NuSTAR spectral analysis performed in Section~\ref{sec:spec} allows the temperatures and emission measures of these quiet Sun impulsive events to be put into the context of previous studies of large flares and microflares from active regions, as well as quiet Sun flares. In Fig.~\ref{fig:templot}, we show the isothermal fitting results of the NuSTAR spectra for the events from this study in comparison to results from previous X-ray studies of flares as EM vs T. We also include in this plot the results for flaring X-ray bright points observed during solar minimum with NuSTAR \citep{paterson2024}, from the same observing campaigns used in this work, as well as the previous NuSTAR quiet Sun flares \citep{kuhar2018}.

In Fig.~\ref{fig:templot}, it can be seen that the events investigated in this paper are much cooler and fainter compared to the previously studied flares and microflares. The temperatures of the quiet Sun impulsive events was found to lie mostly in the narrow range of 3.2--4.0~MK, which is a similar temperature range to that found for the NuSTAR quiet Sun flares studied by \citet{kuhar2018}. However, some of the seven events here had emission measures over an order of magnitude fainter than the quiet Sun flares of \citet{kuhar2018}. This is a consequence of the lower background and NuSTAR's higher livetime at solar minimum allowing fainter events to be detected. The only outlier here is the possible hotter emission seen from Event 2, the mini-filament eruption, shown with the square symbol in Fig.~\ref{fig:templot}.

Some of the quiet Sun flares detected in SXR with XSM \citep{2021ApJ...912L..13V} do overlap in temperature with the NuSTAR events presented in this paper, but are up to several orders of magnitude brighter in emission measure. Several of these XSM flares are hotter, and more overlap with small active region microflares detected previously with NuSTAR -- though these XSM observations were from periods of low solar activity, with no NOAA active regions present.

Overall there is a general scaling that the larger the flare, the higher temperature and bigger emission measure. This has been seen for active region flares on the Sun through to stellar flares, scaling as approximately log emission measure proportional to log temperature \citep{2002ApJ...577..422S,2008ApJ...672..659A}. For active region flares, including smaller microflares, is was found that the scaling flattens at the lower temperatures, being roughly consistent with $EM \propto 10^{aT+b}$, where $a,b$ are some constants. This and subsequent work \citep{hannah2008, 2012ApJS..202...11R} have shown a very large spread in values, particularly for microflares. This wide spread in values can be seen in Fig.~\ref{fig:templot}, which gets even worse for the flares outwith active regions. These flares seem to have a higher emission measure for their temperatures than expected from the scaling for larger flares. This might be indicative that most of the quiet Sun events being a separate population from the active regions flares, the energy release process possibly operating through a different physical mechanism, for instance more into heating than particle acceleration. For comparison, we have speculatively added the higher 5~MK emission fitted to Event 2, shown with the square symbol in Fig.~\ref{fig:templot}. This seems to scale closer to the active regions flares, which could hint at a similarity to the processes producing the hottest emission in mini-filament eruptions. Overall however, there is a substantial spread over several orders of magnitude, particularly with the weakest microflares and quiet Sun flares, resulting in a poor correlation, and hence making it difficult to draw clear conclusions. 

An automated study of NuSTAR flares \citep{2025Masek}, finds similar results with the quiet Sun events being cooler, yet brighter, than expected compared to active region microflares. Event 2 and 4 from the work presented in this paper were found independently in the automated detection approach used by \citet{2025Masek} (as well as Flare 1 from \citep{paterson2024}) and in both cases an isothermal model fit the spectra well, with parameters similar to those given in Table~\ref{tab:fitresults}.

There are however several caveats that must be considered when interpreting the EM vs T plot, shown in Fig.~\ref{fig:templot}. For example, this plot combines X-ray spectral fitting results from several different instruments, all with different temperature sensitivities. In addition, these fits may not all have come from the same point in the events' evolution. Typically for the smallest events a single spectral fit over the whole event is performed, and this can produce a lower temperatures and higher emission measure than fits done during the impulsive start, which is the standard approach for larger events. Spectral fits over the initial phase, instead of the whole time range, for some of the events shown in this paper does produce slightly larger temperatures (and lower emissions) but not enough to significantly shift them into the scaling expected from active region microflares. Another factor to consider is that the flares' spectra were not all fitted with the same type of model. While the quiet Sun events were fitted with isothermal models, the spectra of  flares and microflares are likely to have multiple model components (i.e. thermal and non-thermal), with only the main thermal parameters being plotted in Fig.~\ref{fig:templot}. As can be seen from Section~\ref{sec:dem}, even these small quiet Sun events demonstrate multi-thermal emission, and so the isothermal model used in the spectral fitting, and shown in Fig.~\ref{fig:templot}, is merely a proxy to this underlying more complicated heating.  Taking into account these caveats, in addition to the small sample of NuSTAR quiet Sun flares, it is difficult to make a definitive conclusion but it is suggestive that there might be a difference to the nature of the quiet Sun events compared to the active regions flares.

\section{Conclusions}
\label{sec:conc}

In this paper, we have investigated seven transient X-ray brightenings observed in the quiet Sun during solar minimum with NuSTAR. This included quiet Sun flares, as well as a mini-filament eruption. 

We fitted the NuSTAR spectra for the seven quiet Sun events, finding temperatures in the narrow range between 3.1--4.0~MK, with corresponding emission measures ranging from (0.75--17.0) $\times 10^{43}$~\pcm. We found that almost all of these spectra were well represented by an isothermal model, with no hotter or non-thermal components. The exception was the mini-filament eruption, which (unlike the other events) produced a signature in the AIA Fe XVIII proxy channel. This is indicative of plasma heated to temperatures > 4~MK, and NuSTAR spectrum for this event does show an excess of counts at higher energies but a double thermal fit (adding a 5.1~MK) component did not significantly improve the fit. The DEM analysis of these events confirms the presence of hotter material in this event. 

The temperatures found in this analysis lie in the same range as those found in the previous NuSTAR quiet Sun flare analysis of \citet{kuhar2018}. However, the emission measures found for the events in this study were generally smaller (by more than an order of magnitude for some events) due to the lower background and higher instrument livetime at solar minimum. From the spectral fitting results, we also calculated the thermal energies of the events, finding that they ranged from (2.5--8.9) $\times 10^{25}$~erg. This is an roughly order of magnitude lower than the thermal energies of the events studied by \citet{kuhar2018}, which were between (2--6) $\times 10^{26}$~erg. The temperature of the events studied here have some overlap with quiet Sun flares previously observed in SXR \citep{2021ApJ...912L..13V}, but are consistently hotter than several of EUV brightenings \citep{2023A&A...671A..64D,2024A&A...688A..77D}.

When the results for the seven impulsive events were plotted in as EM vs T together with results from X-ray studies of flares and microflares, we found that the quiet Sun events (from this study and from \citet{kuhar2018} and \citet{paterson2024}) were cooler compared to active region events. By considering the new events studied in this paper in addition to earlier quiet Sun flare results, it can be seen that the scaling of quiet Sun events seems to change compared to the scaling between large flares and microflares. In general, the quiet Sun flares are observed to be slightly offset from this scaling, having higher emission measures or lower temperatures than expected. This may be indicative of quiet Sun flares being a separate population from the larger active region flares and microflares, with the energy release process operating in a different way in which heating more dominates over particle acceleration. However, as was discussed in Section~\ref{sec:tem}, there are several caveats that should be taken into account in the interpretation of the EM vs T plot, particularly the large spread and weak correlation, which make it difficult to make a definitive conclusion.

NuSTAR has allowed the observation of the smallest events in the quiet Sun to be detected in HXRs. However, NuSTAR's limited throughput, combined with faint, short duration events, makes it difficult to detect hotter or non-thermal components in their emission that would give further insight into the underlying physical processes. In the future, a dedicated solar HXR instrument with improved sensitivity and throughput would allow a more rigorous investigation of the scaling between the temperatures and emission measures of larger flares and quiet Sun flares.

\section*{Acknowledgements}

This paper made use of data from the NuSTAR mission, a project led by the California Institute of Technology, managed by the Jet Propulsion Laboratory, funded by the National Aeronautics and Space Administration. We thank the NuSTAR Operations, Software and Calibration teams for support with the execution and analysis of these observations. This research made use of the NuSTAR Data Analysis Software (NUSTARDAS) jointly developed by the ASI Science Data Center (ASDC, Italy), and the California Institute of Technology (USA). Hinode is a Japanese mission developed and launched by ISAS/JAXA, with NAOJ as domestic partner and NASA and UKSA as international partners. It is operated by these agencies in co-operation with ESA and NSC (Norway).  AIA on the Solar Dynamics Observatory is part of NASA’s Living with a Star program. This research has made use of SunPy, an open-source and free community-developed solar data analysis package written in Python \citep{sunpy_community2020}.  This research made use of Astropy, a community-developed core Python package for Astronomy \citep{2018AJ....156..123A, 2013A&A...558A..33A,2022ApJ...935..167A}. SP acknowledges support from the UK’s Science and Technology Facilities Council (STFC) doctoral training grant (ST/T506102/1). IGH acknowledges support from a Royal Society University Fellowship (URF/R/180010) and STFC grants (ST/T000422/1, ST/X000990/1).  

\section*{Data Availability}

All the data used in this paper are publicly available. In particular, NuSTAR via the  \href{https://heasarc.gsfc.nasa.gov/db-perl/W3Browse/w3table.pl?tablehead=name=numaster&Action=More+Options}{NuSTAR Master Catalog} with the OBSIDs 80512218001--80512228001 for the 21 February 2020 observation, and 80600201001 and 80610202001--80610210001 for 12--13 September 2020.

\bibliographystyle{mnras}
\bibliography{example} 





\bsp	
\label{lastpage}
\end{document}